\documentclass[fleqn,usenatbib]{mnras}
\usepackage[T1]{fontenc}
\DeclareRobustCommand{\VAN}[3]{#2}
\let\VANthebibliography\thebibliography
\def\thebibliography{\DeclareRobustCommand{\VAN}[3]{##3}\VANthebibliography}

\usepackage{graphicx}	
\usepackage{amsmath}	

\usepackage{epsfig}
\usepackage{epstopdf}

\usepackage{newtxtext,newtxmath}

\title[]{Modulation of the solar microwave emission by sausage oscillations}

\author[E. G. Kupriyanova et al.]{
Elena G. Kupriyanova,$^{1}$\thanks{E-mail: elenku@bk.ru (EGK)}
Tatyana I. Kaltman,$^{2}$
and Alexey A. Kuznetsov$^{3}$
\\
$^{1}$Central Astronomical Observatory at Pulkovo of RAS, Saint Petersburg, 196140,  Russia\\
$^{2}$Saint Petersburg Branch, Special Astrophysical Observatory of RAS, Saint Petersburg, 196140, Russia\\
$^{3}${Institute of Solar-Terrestrial Physics SB RAS, Irkutsk, 664033, Russia}
}

\date{Accepted XXX. Received YYY; in original form ZZZ}

\pubyear{2022}

\begin{document}
\label{firstpage}
\pagerange{\pageref{firstpage}--\pageref{lastpage}}
\maketitle

\begin{abstract}
The modulation of the microwave emission intensity from a flaring loop by a standing linear sausage fast magnetoacoustic wave is considered in terms of a straight plasma slab with the perpendicular Epstein profile of the plasma density, penetrated by a magnetic field. The emission is of the gyrosynchrotron (GS) nature, and is caused by mildly relativistic electrons which occupy a layer in the oscillating slab, i.e., the emitting and oscillating volumes do not coincide. It is shown that the microwave response to the linear sausage wave is highly non-linear. The degree of the non-linearity, defined as a ratio of the Fourier power of the second harmonic to the Fourier power of the principal harmonic, is found to depend on the combination of the width of the GS source and the viewing angle, and is different in the optically thick and optically thin parts of the microwave spectrum. This effect could be considered as a potential tool for diagnostics of the transverse scales of the regions filled in by the accelerated electrons.
\end{abstract}



\section{Introduction}

Solar flares are one of the most powerful phenomena in the solar atmosphere. Physical processes operating in flares at kinetic and magnetohydrodynamic (MHD) scales, such as the charged particle acceleration and magnetic reconnection, are fundamental processes of plasma astrophysics and remain subject to intensive studies. An important method for probing the plasma in flaring sites is MHD seismology, based upon oscillatory processes detected in flares \citep[e.g.][]{2020STP.....6a...3K, 2021SSRv..217...66Z}. Among the oscillations which are most commonly observed in flaring regions are standing fast magnetoacoustic waves of the sausage symmetry, i.e. the axisymmetric oscillatory motions across the magnetic field, characterised by variations of the plasma density and absolute value of the magnetic field \citep[see, e.g.,][for recent comprehensive reviews]{2020SSRv..216..136L}. As the typical periods of sausage oscillations are from a few seconds to several tens of seconds \citep[e.g.,][]{2009SSRv..149..119N}, their detection requires the use of instruments with high time resolution. In particular, sausage oscillations are often observed as the modulation of coherent and non-coherent radio and microwave emissions.  For example, sausage oscillations have been used for the interpretation of the time variations of the gyrosynchrotron (GS) emission intensity from spatially resolved coronal loops \citet{2003A&A...412L...7N, 2005A&A...439..727M}, and simultaneous oscillations of GS and soft X-ray emission along a loop \citep{2008A&A...487.1147I}, the wiggling of zebra-pattern lanes \citep{2013ApJ...777..159Y}, and the precipitation rate of the nonthermal electrons at the opposite footpoints of a loop \citep{2018ApJ...859..154N}.  

Sausage oscillations are used as seismological probes of perpendicular profiles of the density and the absolute value and twisting of the magnetic field in flaring plasma structures, such as plasma loops \citep[e.g.,][]{2020SSRv..216..136L}. Both regular and random perpendicular profiles could be addressed \citep[e.g.][]{2007SoPh..246..165P, 2014A&A...567A..24H, 2015ApJ...812...22C, 2015ApJ...810...87L}. Typically, the analysed plasma structures are modelled as a plasma slab or a cylinder, surrounded by a plasma with different properties \citep[see, e.g.][]{2012ApJ...761..134N, 2014A&A...567A..24H}, respectively. The consideration of sausage oscillations in a loop with a varying cross-section demonstrated that the slab or cylinder models are adequate. For example, for the fundamental parallel harmonic, the cross-section radius increase near the loop top by a factor of 2 causes the decrease in the oscillation period of about 5\% only \citep{2009A&A...494.1119P}.

Despite intensive theoretical and observational studies of sausage oscillations, the link of the theory and observational outcomes remains non-trivial. One of the difficulties is the effect of the line-of-sight (LoS) integration in the optically thin regime. Let us illustrate it for the case of the EUV or soft X-ray emissions. On one hand, a sausage oscillation is characterised by the perturbations of the plasma density. One the other hand, if the LoS is perpendicular to the oscillating plasma structure, the plasma is displaced along the LoS. As the observed emission intensity is proportional to an integral of the squared density along the LoS, the observed intensity does not change much in different phases of the oscillation. This effect was first demonstrated by \citet{2012A&A...543A..12G}, and then confirmed by comprehensive forward modelling \citep{2013A&A...555A..74A}. A similar problem appears in the radio observations \citep[see, e.g.][]{2012ApJ...748..140M}, and has been addressed by forward modelling of microwave observables by \citet{2015A&A...575A..47R, 2015SoPh..290.1173K}.  

In contrast with the modelling of the thermal emission which is determined by the variation of the plasma density along the LoS, various kinds of non-thermal radio emissions are also determined by the magnetic field non-thermal electrons, LoS angle to the field, and their distributions along the LoS. In the majority of forward modelling studies, the spatial structure of non-thermal electrons has been considered to coincide with the oscillatory perturbations of the plasma density and magnetic field by a sausage wave \citet{2015A&A...575A..47R, 2015SoPh..290.1173K}. However, this assumption could be incorrect, as the volumes occupied by non-thermal emission and involved in the sausage oscillation are determined by different physical processes.
The perpendicular structure of the sausage wave perturbation is determined by parameters of wave-guiding plasma non-uniformity \citep[e.g.,][]{2020SSRv..216..136L}. The magnetic flux tube filled in with non-thermal electrons is determined by the acceleration mechanism. For example, a coronal structure, which is observed in microwaves as a single loop, appears as a filamented structure in EUV \citep[e.g.,][]{2013AstL...39..267Z}. This structure can oscillate as a whole, being perturbed by a sausage wave, while only one its filaments may contain energetic electrons emitting the microwaves. 

The mismatch of the oscillating and emitting volumes could provide us with important seismological constrains on the enigmatic non-thermal electron acceleration processes. The microwave emission could be considered as a tool for diagnostics of the low-amplitude perturbation of the plasma magnetic by the sausage wave because of the high sensitivity and the non-linear dependence of the intensity of the microwave emission on the both magnitude and direction of the magnetic field and on the parameters of the emitting electrons \citep[e.g.,][]{1982ApJ...259..350D}, resulting in the non-linear microwave response to the linear sausage wave. 
The aim of this paper is to study how the degree of this non-linearity depends on both the width of the emitting volume and the LoS angle to the magnetic field.

The paper is organised as follows. The model and basic equations are described in Section~\ref{s:MHD}. The source of the microwave emission and details of the calculation of the microwave emission are presented in Section~\ref{s:radio}, and details of obtaining the microwave light curves are presented in Section~\ref{s:LC}. Section~\ref{s:results} presents the principal results of the simulations, which are summarized and discussed in Section~\ref{s:discussion}.

\section{Plasma slab and MHD wave} \label{s:MHD}
We consider a transversely inhomogeneous (along the $x$-axis) plasma slab stretched along a uniform magnetic field $B_0$ directed along $z$-axis. The slab is perturbed by a symmetric (sausage) fast magnetoacoustic MHD mode of the slab. The system is uniform in the $y$ direction. If we place an infinitely remote observer in the $xz$ plane, then we may ignore the $y$ direction and simulate the MHD wave in the two-dimensional slab. So, the components of the unperturbed magnetic field are $B_{z0} = B_0 = \textrm{const}$, $B_{x0} = 0$.

\begin{figure*}
	\centering{
		\includegraphics[width=1.0\textwidth]{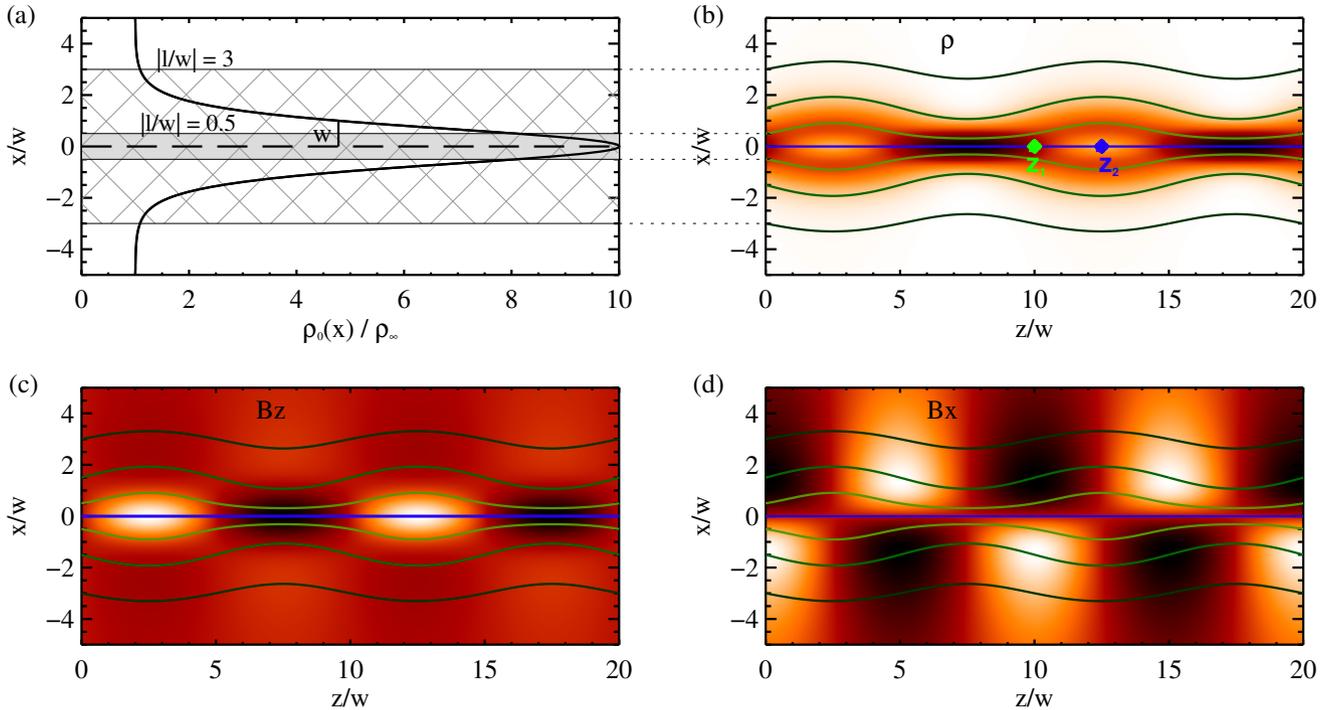}
	}
	\caption{ Panel (a): normalised transverse profile of the non-perturbed thermal plasma density. The vertical short line indicates the characteristic slab width $w$. Two rectangles illustrate homogeneous distributions of the density of the accelerated electrons within both the narrow GS source ($|l/w| = 0.5$, grey area) and the wide GS source ($|l/w| = 3$, cross-hatched area). Panel (b): perturbed plasma density $\rho$. Panel (c): perturbed  $B_{z}$ component of the magnetic field. Panel (d): perturbed  $B_{x}$ component of the magnetic field. The darkest colour corresponds to the maximal value, while the lightest colour~--- to the minimal value. The green point $z_1$ in panel (b) denotes one of the sausage wave nodes and the blue point $z_2$~--- one of the anti-nodes. Different pairs of the same-coloured wavy lines delimit the GS sources having different widths.}
	\label{f:rho_Bx_Bz}
\end{figure*}

The transverse profile of the plasma non-uniformity, shown by black curve in panel (a) in Figure~\ref{f:rho_Bx_Bz}, is defined by the Epstein function \citep{2003A&A...409..325C}
\begin{equation}\label{eq:Epstein}
	\rho_0(x) = \rho_\mathrm{max} \rm{sech}^2 \it \frac{x}{w} + \rho_{\infty}
\end{equation}
with the characteristic width $w$ referred to as the slab width, and the density contrast 
\begin{equation}\label{eq:dc}
	d=\frac{\rho_\mathrm{\max}+\rho_{\infty}}{\rho_{\infty}}.
\end{equation}
The system is symmetric relatively to $z$ axis (blue line in Figure~\ref{f:rho_Bx_Bz}) where the plasma density is $\rho_\mathrm{\max}+\rho_{\infty}=\rho_0(x=0)$. Plasma density at the point infinitely remote from the $z$ axis is $\rho_{\infty}=\rho_0(x\to\infty)$. 

Under conditions typical of the solar corona, where the plasma parameter $\beta \ll 1$, the ideal MHD equations \citep[e.g.,][]{1997JPlPh..58..315N} can be linearised, and the perturbed plasma density and the perturbed magnetic field components can be expressed via the transverse component of the perturbed speed $\tilde{V}_{x} = \tilde{V}_x(x,z,t)$:
\begin{equation}\label{eq:rho}
	\tilde{\rho} = - \int \frac{\partial (\rho_0 \tilde{V_{x}})}{\partial x} dt, 
\end{equation}
\vspace{-4mm}
\begin{equation}\label{eq:Bx}
	\tilde{B_{x}} = B_0 \int \frac{\partial \tilde{V_{x}}}{\partial z} dt, \:\:\:\:\:\:\:\:\:\:\:\:
	\tilde{B_{z}} = - B_0 \int \frac{\partial \tilde{V_{x}}}{\partial x} dt. 
\end{equation}
Since we consider the standing sausage wave, which is periodic in the $z$ direction with the longitudinal wave number $k$ and wavelength $\lambda = 2\pi/k$, the general expression of the transverse component $\tilde{V}_{x}$ \citep[see, e.g.,][]{1995SoPh..159..399N} can be simplified:
\begin{equation}\label{eq:Vx}
	\tilde{V}_{x}(x,z,t) = A U(x) \sin(kz)\cos(k V_\mathrm{ph} t).
\end{equation}
Here $A$ is the amplitude of the MHD perturbation normalized by the Alfv\'en speed $C_\mathrm{A0}$ at the slab axis ($x=0$), $A \ll 1$, and $V_\mathrm{ph} = \omega / k$ is the phase speed. Analytical expression for the transverse structure $U(x)$ of the sausage wave
\begin{equation}\nonumber
	U(x) = \frac{ \sinh(x/w) }{ \cosh^{\zeta}(x/w)}, \:\:\:\:\:\:\:\:\:
	\zeta = \frac{ |k|w }{ C_\mathrm{A \infty }} \sqrt{ C_\mathrm{A \infty}^2 - V_\mathrm{ph}^2} +1,
\end{equation}
for the plasma density inhomogeneity defined by Equation~(\ref{eq:Epstein}) was obtained by \citet{2003A&A...409..325C}. The phase speed is defined by the dispersion equation 
$$
\frac{ |k|w }{ C_\mathrm{A0}^2} (V_\mathrm{ph}^2 - C_\mathrm{A0}^2) - \frac{2}{|k|w} = \frac{3}{C_\mathrm{A \infty}} \sqrt{C_\mathrm{A \infty}^2 - V_\mathrm{ph}^2},
$$
here $C_\mathrm{A \infty}$ is the Alfv{\'e}n speed at the infinitely remote point ($x \to \infty$).

Following Equations (\ref{eq:rho})--(\ref{eq:Vx}), the perturbed values for plasma density and the components of the magnetic field become
\begin{equation}
	\rho = \rho_0 + \tilde{\rho}, \:\:\:\:\:\:\:\:\:  
	B_{z} = B_0 + \tilde{B}_{z}, \:\:\:\:\:\:\:\:\:  
	B_{x} = \tilde{B}_{x}.
	\label{eq:RhoB_Total}
\end{equation}
Distributions of these parameters in the $xz$ plane at the moment of the maximal perturbation are shown in panels (b), (c), (d) in Figure~\ref{f:rho_Bx_Bz} where axes $x$ and $z$ are normalised by the characteristic width of the plasma non-uniformity $w$. In this paper, for simplicity, we consider strictly monochromatic (i.e. strictly
repetitive) decayless MHD wave; therefore, it is sufficient to consider one oscillation period. The wavelength of the MHD wave was chosen to be
$\lambda=10w$.

\section{Microwave emission source} \label{s:radio}
Thermal plasma in the slab is the source of thermal free-free radio emission. In addition, if there is a population of electrons accelerated up to mildly relativistic energies, such a system is the source of gyrosynchrotron emission, which is usually associated with solar radio emission in the microwave band. The GS source is populated with energetic electrons isotropically distributed over pitch-angles and power-law distributed over energy; the concentration of the electrons is assumed to be constant in the unperturbed source which, according to the above-described approach, occupies a part of the sausage-oscillating slab. The GS source is enclosed between pairs of the magnetic field lines symmetric relatively to the $z$ axis. In the unperturbed slab, the magnetic field lines are straight lines at the distances of $x/w=l/w$ from the $z$ axis; thus $|l/w|$ is the half-width of the GS source. We consider the GS sources both narrower ($|l/w| \leq 1$) and wider ($|l/w| > 1$) than the characteristic slab width. Examples of the unperturbed electron density distributions are shown in Figure~\ref{f:rho_Bx_Bz} (panel (a)) by both grey rectangle for the narrow GS source ($|l/w| = 0.5$) and the cross-hatched rectangle for the wide GS source ($|l/w| = 3$). Hereinafter, we will omit the modulus sign from the width of the GS source, for the simplicity.

The sausage wave perturbs the parameters of the slab and, therefore, modifies the shape of the magnetic field lines and, thereby, curves the boundaries of the GS source along the slab. Examples of the pairs of the symmetric perturbed magnetic field lines, which are the outer boundaries of the GS sources, are shown in Figure~\ref{f:rho_Bx_Bz} (panels (b), (c), (d)) by the same-coloured wavy curves. The local density of the non-thermal particles was assumed to be inversely proportional to the local width of the GS source, to provide that the flux of the accelerated electrons through the cross-section of the slab is conserved. So, the 
modulation of the local distance between two symmetric magnetic field lines leads to the modulation of the electron density along $z$ axis. Modulation of the energy and pitch-angle distributions by the MHD oscillations was neglected. 

We use the Fast Gyrosynchrotron Codes \citep{Fleishman2010, Kuznetsov2021} to calculate the radio emission from the slab. As an input data, these codes use the specified distributions of the emission source parameters  (such as the thermal plasma density and temperature, concentration and distribution characteristics of the nonthermal electrons, and the magnetic field strength and direction) along a line-of-sight. The codes include the contributions of both the GS (nonthermal) and free-free (thermal) emission mechanisms. However, we note that the thermal emission is several orders of magnitude weaker than the GS emission, for the model parameters listed below. The codes calculate the resulting radio emission by integrating the radiation transfer equations for both the left-hand ($L$) and right-hand ($R$) polarised components, which correspond to either ordinary or extraordinary electromagnetic modes depending on the local magnetic field direction.  

We consider the following parameters of the unperturbed slab: magnetic field strength $B_0=200$ G, plasma temperature $T_0=10^7$~K, plasma density at the $z$ axis $\rho_\mathrm{max} = 5 \times 10^9$~cm$^{-3}$, plasma density outside the slab $\rho_\infty = \rho_\mathrm{max}/(d-1)$, where the density contrast $d = 10$ corresponds to the plasma in flaring loops, the characteristic slab width $w=3\times 10^8$~cm. The nonthermal electrons in the unperturbed GS source are assumed to have the constant concentration of $n_{\mathrm{b}}=10^5$ $\textrm{cm}^{-3}$, power-law energy distribution in the range from 0.1 to 10~MeV with the spectral index of $\delta = 3$, and isotropic pitch-angle distribution. These parameters were chosen to provide the GS emission spectral peak at around 3--5~GHz, which is a typical value in solar flares. We consider the $z$ axis inclined relatively to the line-of-sight by the viewing angle $\theta$ from 40$^\circ$ to 89$^\circ$.

\section{Light curves} \label{s:LC}
Intensity of the GS emission strongly and non-linearly depends on both the local magnitude and direction of the magnetic field (see, for example, Equation~(1) in \citep{Fleishman2010} or the empirical approximation in \citep{1982ApJ...259..350D}). Standing sausage wave modulates parameters of the GS source which results in oscillations of the observed radio flux. 

Application of the Fast GS Codes allows one to obtain the distribution of the intensity ($I = R+L$) of the radio emission along the plane-of-sky which is inclined relatively to the $z$ axis by the angle $90^\circ - \theta$. Running the procedure for each time during the period of the sausage wave, one obtains the temporal evolution of the intensity (light curve). We split the period of the wave into 20 time segments, so, by definition, the period of the sausage wave $P_\mathrm{saus} = 20$ time units (t.\,u.).  We consider the light curves at two positions: $z_1$ which is a node of the sausage wave and $z_2$ which is an anti-node (see panel (b) in Figure~\ref{f:rho_Bx_Bz}). A few representative examples of the light curves, smoothed with the cubic spline interpolation, are presented in Figures~\ref{f:timeprofs_17_50}--\ref{f:timeprofs_03_50} (upper row in each figure). The light curves are coloured in green for $z_1$ and blue for $z_2$. We consider both the optically thin emission at 17~GHz (Figures~\ref{f:timeprofs_17_50}--\ref{f:timeprofs_17_70}) and the optically thick emission at 1.5~GHz (Figure~\ref{f:timeprofs_03_50}). In each of these figures, the upper left panel corresponds to a narrower GS source ($l/w = 0.1$), while the upper right panel --- to a wider GS source ($l/w = 3$ or $l/w = 1.5$). Figures~\ref{f:timeprofs_17_50} and Figure~\ref{f:timeprofs_03_50} correspond to $\theta = 50^\circ$ while Figure~\ref{f:timeprofs_17_70}~--- to $\theta = 70^\circ$. 

\section{Results} \label{s:results}
\begin{figure*}
	\centering{
		\includegraphics[width=0.8\textwidth]{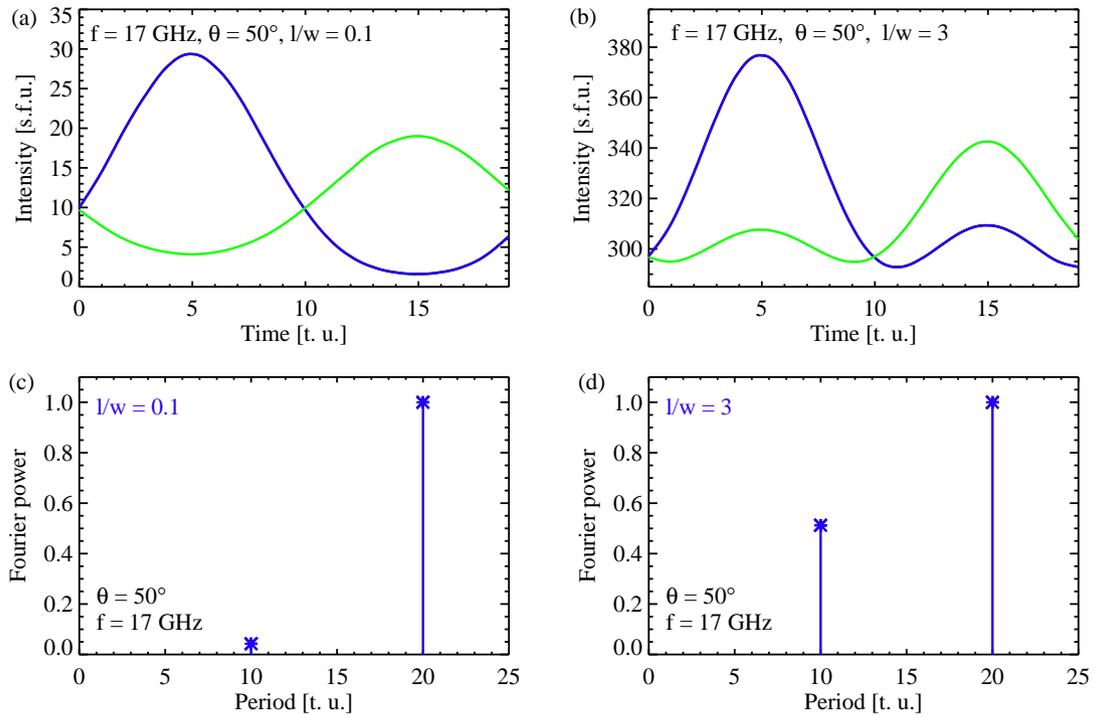}
	}
	\caption{The upper row: the microwave light curves in the optically thin regime ($f = 17$~GHz) from the sausage-oscillating slab, both for the narrower GS source, $l/w = 0.1$ (panel (a)) and for the wider GS source, $l/w = 3$ (panel (b)).  In each panel, the green and blue curves are obtained along line-of-sight passing through the node $z_1$ and the anti-node $z_2$, respectively (see panel (b) in Figure~\ref{f:rho_Bx_Bz}) at the viewing angle $\theta = 50^\circ$. The lower row: the normalized Fourier powers (asterisks) of the periods $P_1$ and $P_2$ found in the light curve associated with the anti-node (see blue curves in the upper panels) for both the narrower GS source (panel (c)) and the wider GS source (panel (d)).}
	\label{f:timeprofs_17_50}
\end{figure*}
\begin{figure*}
	\centering{
		\includegraphics[width=0.8\textwidth]{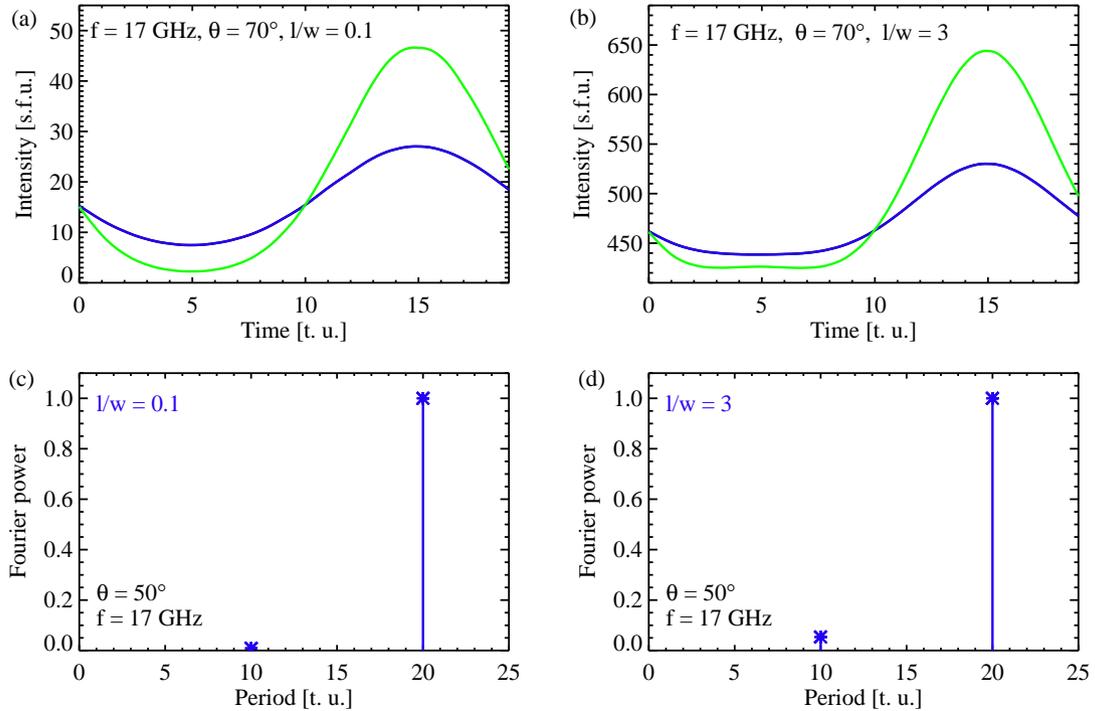}
	}
	\caption{Same as in Figure \protect\ref{f:timeprofs_17_50}, for the optically thin regime ($f=17$ GHz) and viewing angle of $\theta=70^{\circ}$.} \label{f:timeprofs_17_70}
\end{figure*}

\begin{figure*}
	\centering{
		\includegraphics[width=0.8\textwidth]{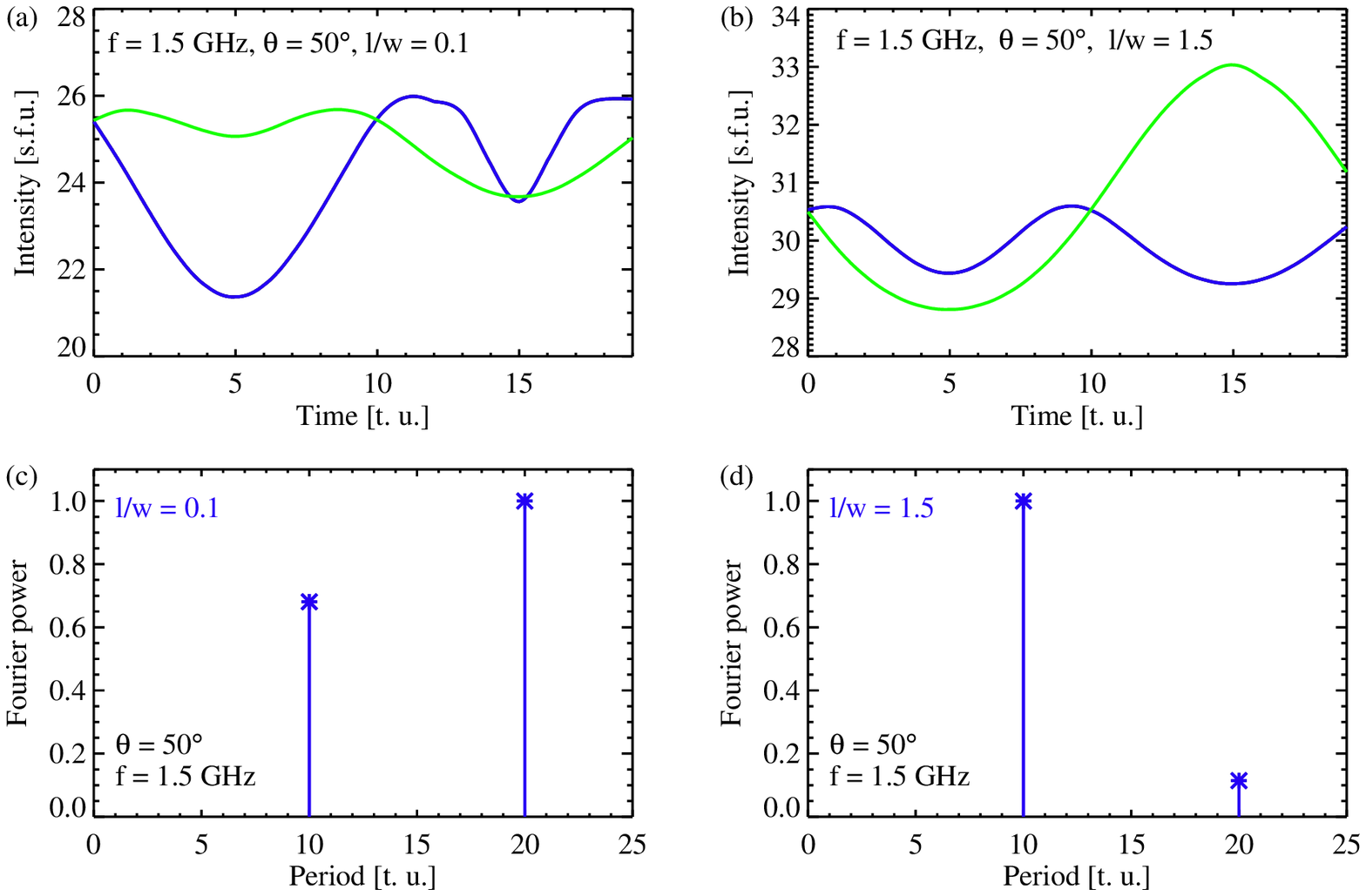}
	}
	\caption{Same as in Figures \protect\ref{f:timeprofs_17_50}--\protect\ref{f:timeprofs_17_70}, for the optically thick regime ($f=1.5$\,GHz) and viewing angle of $\theta=50^{\circ}$.}
	\label{f:timeprofs_03_50}
\end{figure*}

It is clearly seen in Figures~\ref{f:timeprofs_17_50}--\ref{f:timeprofs_03_50} that the light curves deviate from a sinusoidal shape, i.e., the response of the radio emission to the linear MHD perturbation is nonlinear. Moreover, this deviation depends on the width of the GS source $l/w$ and the viewing angle $\theta$. For example, for $\theta = 50^\circ$, the non-linearity is more pronounced for the narrower GS source, which is manifested as a flattening of minima or maxima of the light curves (e.g., blue curve in panel (a) of Figure~\ref{f:timeprofs_17_50}). In contrast, for $\theta = 70^\circ$, the flattening and hence non-linearity is more pronounced for the wider GS source (see both blue and green curves in panel (b) of Figure~\ref{f:timeprofs_17_70}). 

\begin{figure*}
	\centering{
		\includegraphics[width=0.8\textwidth]{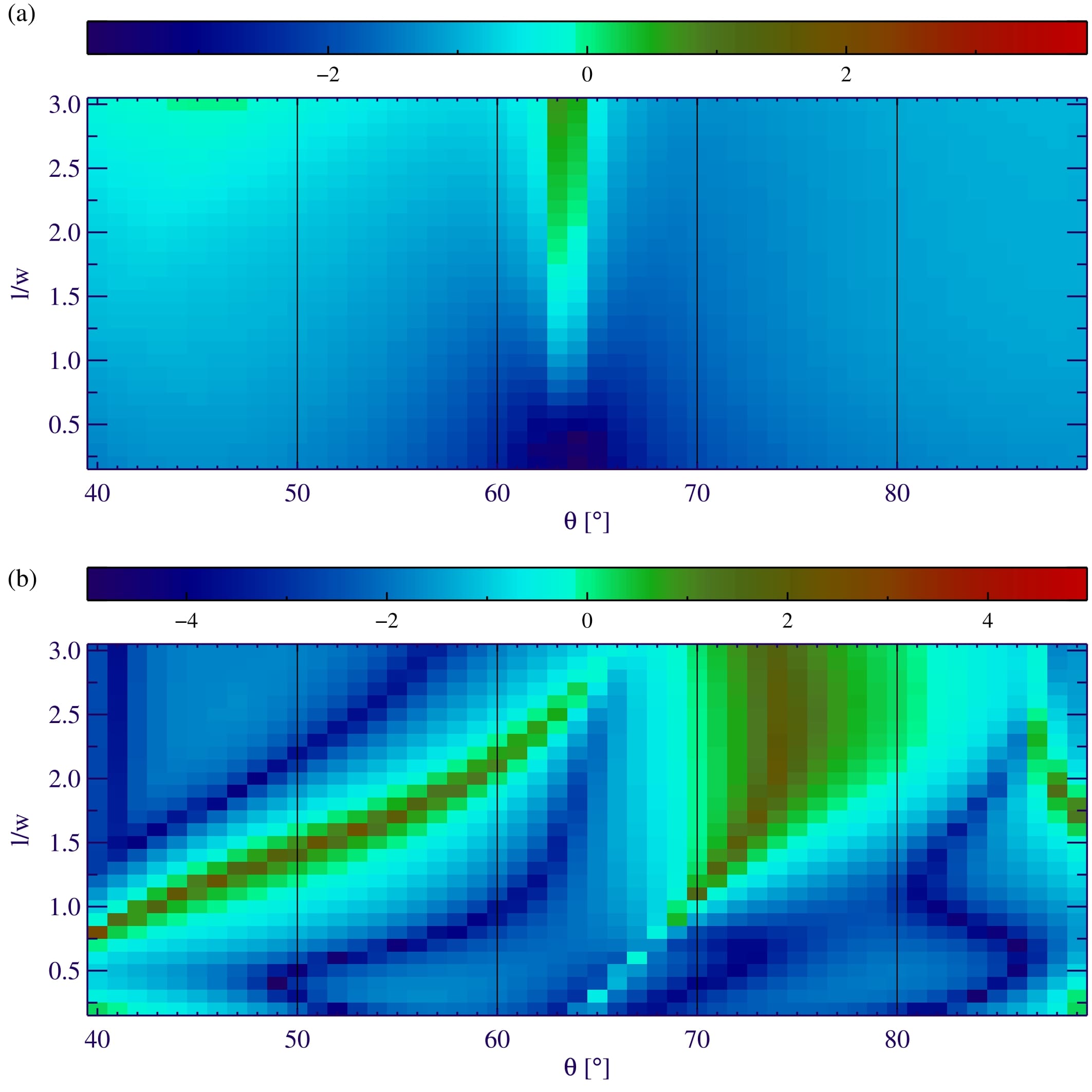}
		\includegraphics[width=0.8\textwidth]{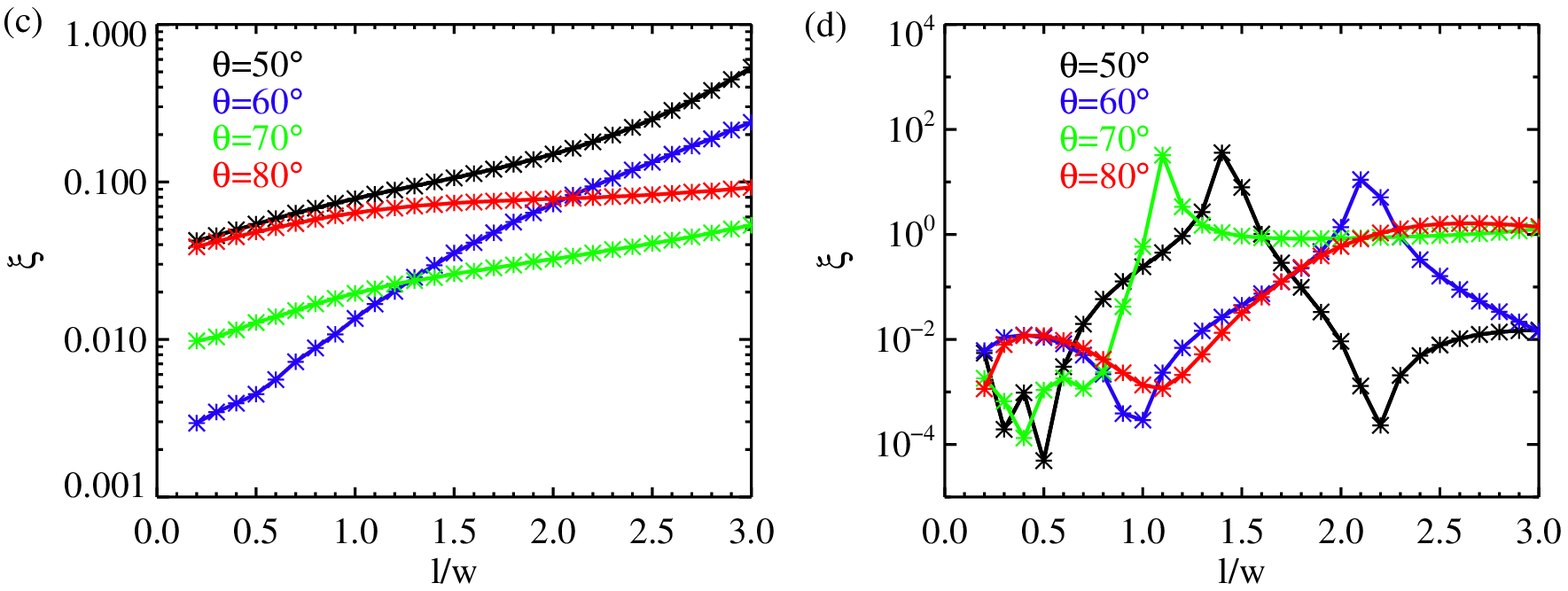}
	}
	\caption{Distributions of the ratios ($\log_{10} \xi$) of the Fourier power of the second harmonic with the period $P_2 = 10$~t.\,u. to the Fourier power of the principal harmonic with the period $P_1 = 20$~t.\,u. relatively to the width of the GS source ($l/w$) and viewing angles $\theta$ in both the optically thin regime at 17\,GHz (panel (a)) and the optically thick regime at 1.5\,GHz (panel (b)). The lighter colours correspond to the higher values of $\log_{10} \xi$. 
		Lower row: the ratios $\xi$ for the selected viewing angles $\theta$, marked by vertical lines in the upper panels, for optically thin regime (panel (c)) and for optically thick regime (panel (d)). }
	\label{f:Fourier_ratios}
\end{figure*}

To estimate the degree of the non-linearity, we calculated the modulation depth 
$$
\Delta = \frac{I_{max} - I_{min}}{I_{mean}},
$$
where $I_{max}$ and $I_{min}$ are the maximal and minimal intensities during the MHD wave period, and $I_{mean}$ is the respective mean value. For the narrower GS source, the modulation depths are of $\Delta \sim 100$--200\% and $\Delta \sim 60$--130\% in the optically thin and optically thick regimes, respectively. The high values of $\Delta$ are comparable with those usually associated with  quasi-periodic ejection of electrons \citep[e.g.,][]{2016SoPh..291.3427K}. 
The deep modulation can be explained by the location of the narrow GS source near the $z$ axis where the oscillations magnitude of the waveguide parameters is maximal.

For the wider GS source, the modulation depth is lower: $\Delta \sim 10$--20\%  for both the optically thin and optically thick emissions. On the other hand, the non-linear effect is much stronger: an extra peak or dip in the lightcurves appears either at $t = P_\mathrm{saus}/4 = 5$ (as in the green curves in panel (b) of Figure~\ref{f:timeprofs_17_50} and panel (a) of Figure~\ref{f:timeprofs_03_50} or as in the blue curve in panel (b) of Figure~\ref{f:timeprofs_03_50}) or at $t = 3P_\mathrm{saus}/4 = 15$ (as in the blue curves in panel (b) of Figure~\ref{f:timeprofs_17_50} and panel (a) of Figure~\ref{f:timeprofs_03_50}). 
The amplitude of the extra peak/dip varies from around 10\% to 100\% of the above-mentioned modulation depth $\Delta $ for different combinations of $l/w$ and $\theta$. 

Lower panels in Figures~\ref{f:timeprofs_17_50}--\ref{f:timeprofs_03_50} demonstrate the ideal Fourier periodograms,
where one of the peaks corresponds to the principal period of the sausage wave at $P_1 = P_\mathrm{saus} = 20$~t.\,u. The nonlinear response of the radio emission mechanism to the MHD oscillation results in appearance of the second harmonic with the period $P_2 = 10$~t.\,u.  The principal period dominates in most of the periodograms, while the second harmonic is less intensive. However, for some combinations of $l/w$ and $\theta$, the Fourier power of the second harmonic becomes higher; an example is presented in Figure~\ref{f:timeprofs_03_50} (panel (d)).

We should stress that the second harmonic obtained in our study, which the period exactly equals to the half-period of the principal harmonic, $P_2 = 0.5 P_1$, differs from the second harmonic of the eigen sausage mode, which the period is greater than $0.5 P_1$ because of the wave dispersion \citep[e.g.,][]{2005LRSP....2....3N}.

We characterize of the degree of non-linearity as a ratio ($\xi$) of the Fourier power of the second harmonic to that of the principal  harmonic.
Examples of the $\xi$ maps, as functions of the source width and the viewing angle, are shown in Figure~\ref{f:Fourier_ratios} for both the optically thin (panel (a)) and optically thick (panel (b)) regimes. The maps correspond to the anti-node, $z_2$ (see Figure~\ref{f:rho_Bx_Bz}); note that we plot $\log_{10} \xi$ values here. In the optically thin regime, the $\xi$ values increase with the increasing of the width $l/w$ for most angles $\theta$. Besides, there is a bright peculiarity at the viewing angles around $\theta = 63^\circ$, where the second harmonic dominates over the principal  harmonic. In the optically thick regime, the map is more complicated. 

Different characteristic properties of the ratio $\xi$ in both cases are also clearly seen along the cross-sections of the maps at different selected angles $\theta$ ($\theta = 50$, 60, 70, and 80$^\circ$), marked by the vertical lines in two upper panels of Figure~\ref{f:Fourier_ratios}. The cross-sections are presented in panels (c) and (d) of Figure~\ref{f:Fourier_ratios} for the optically thin and optically thick regimes, respectively. 

\section{Summary and discussion} \label{s:discussion}
We performed simulations of the microwave radio response to a standing sausage wave perturbing a zero-$\beta$ plasma slab, and, thereby, the source of the GS emission, modulated by the sausage wave.
The axis of the slab is directed along the magnetic field, and the perpendicular profile of the plasma density profile is smooth. The principal novel and important element which differs this study from the previous ones is that the GS-emitting electrons fill only a layer stretched along the axis of the slab, and localised in the perpendicular direction, i.e., the oscillating and emitting volume do not coincide.  
Our results confirm that the microwave response to a linear MHD wave can be highly non-linear. This result was expected from, for example, simplified expressions for the emission and absorption coefficients \citep{1982ApJ...259..350D} and was found previously in simulated microwave light curves \citep{2012ApJ...748..140M} in a simplified model. The principally new result of our study is that the degree of the non-linearity depends on both the relative width of the GS source inside the oscillating volume, and the viewing angle. The non-linearity appears in the deviation of the microwave intensity light curve from a sinusoidal shape of the sausage oscillation. This results in appearing the second and higher harmonics of the monochromatic sausage oscillation in Fourier spectra of the time-varying GS signal.
Analysis of the degree of non-linearity defined here as a ratio of the Fourier power of the second harmonic to that of the principal harmonic allowed us to define conditions where the non-linear effect is the strongest.

The effect found can be considered as a potential tool for the diagnostics of the transverse scales of the regions containing accelerated electrons within  sausage-oscillating plasma structures.
If we neglect the effect of non-thermal electron diffusion across the magnetic field and take into account the divergence of magnetic tubes in the cusp, this transverse size can be considered as a characteristic perpendicular size of the acceleration region and, therefore, as an upper limit for the perpendicular size of the magnetic reconnection region. Note that previously this size was defined from numerical simulations and was prescribed by certain MHD models. The results presented in our paper open a possibility to estimate the spatial scales of the reconnecting process independently, by analysing the QPPs in the microwave emission.

This independent (observational) knowledge  is important for empirically constraining QPP models involving the reconnection, for example, for coalescence of two magnetic loops
\citep[][]{1987ApJ...321.1031T, 2016PhRvE..93e3205K} or for magnetic tuning fork model \citep[][]{2016ApJ...823..150T}, and also flapping oscillations of the macroscopic current sheet above the loop arcade \citep[][]{2012SoPh..277..283A}. 

In addition, the results obtained are relevant to the interpretation of multi-modal QPPs in the solar \citep[e.g.,][]{2005A&A...439..727M} and stellar \citep[e.g.,][]{1983ApJ...272L..15L} flares, as the observed second, and sometimes third time harmonics can correspond to the same monochromatic MHD mode. However, the difference between the second harmonic obtained in our study from the second harmonic of the eigen sausage mode is that the non-linearity of the microwave response leads to its period equals exactly to the half-period of the principal  harmonic, while for the eigen sausage mode it is not so,  because of the wave dispersion \citep[e.g.,][]{2009A&A...493..259I}.

We should stress that our model is deliberately simple, aiming to demonstrate the discussed effect. A detailed analysis needs to account for the 3D geometry \citep[see, e.g.,][]{2015ApJ...801...23L, 2015ApJ...812...22C, 2015ApJ...810...87L, 2021MNRAS.505.3505K}, the beam size of a radio instrument, variation of the GS spectrum, wave dispersion, etc. An important research avenue is forward modelling of the manifestation of sausage oscillations in the data delivered by the future Square Kilometre Array (SKA) instrument, which is expected to well resolve these oscillations in both time, space and GS spectrum \citep{2015aska.confE.169N}. 

\section*{Acknowledgements}

This study is supported by the Russian Science Foundation project No. 21-12-00195. 

\section*{Data Availability}

The results obtained in the paper are theoretical; no real observations have not been used. Distributions of parameters in the magneto-plasma slab were obtained using equations in Section~\ref{s:MHD}, and their digital version is available on request to the corresponding author.



\bsp	
\label{lastpage}
\end{document}